\documentclass{article}

\usepackage{natbib}
\usepackage{graphicx}

\renewcommand{\vec}[1]{\ensuremath{\mathbf{#1}}}  

\newcommand{\B}{\vec{B}}        
\newcommand{\E}{\vec{E}}        

\newcommand{\GJ}[1]%
{\ensuremath{#1_\textrm{\tiny GJ}}} 

\newcommand{\Rgj}{\GJ{\rho}}       

\newcommand{\RS}{\ensuremath{R_\mathrm{NS}}}    
\newcommand{\RLC}{\ensuremath{R_\mathrm{LC}}}   

\newcommand{\x}{\mathbf{\times}}     
\newcommand{\pd}{\partial}           

\newcommand{\Uvec}[1]%
{ \ensuremath{\mathbf{e}_{#1}} } 

\graphicspath{{./figures/}}

\title{High resolution numerical modeling of the force-free
  pulsar magnetosphere}

\author{Andrey Timokhin\\ 
  Sternberg Astronomical Institute, Moscow, Russia}

\begin{document}
\maketitle

\begin{abstract}
  We consider the force-free magnetosphere of an aligned rotator.  Its
  structure is well described by a Grad-Shafranov equation for
  poloidal magnetic field, the so-called "pulsar equation". We solve
  this equation by a multigrid method with high numerical resolution.
  On the fine numerical grid current layer along the last closed field
  line could be accurately incorporated into the numerical procedure
  and all physical properties of the solution such as Goldreich-Julian
  charge density, drift velocity, energy losses etc.  could be
  accurately calculated. Here we report results of the simulations.
  Among other interesting properties, the solution is not unique, i.e.
  the position of the last closed field line, and hence the pulsar
  energy losses, are not determined by the global magnetosphere
  structure and depend on kinetic of electromagnetic cascades. We
  discuss the properties of the solutions and their implications for
  pulsars. A model for pulsar magnetosphere evolution is proposed. In
  the frame of this model values of the pulsar braking index less than
  3 have natural explanation.
\end{abstract}

\section{Pulsar Equation}
\begin{figure}
  \centering
  \includegraphics[width=\textwidth]{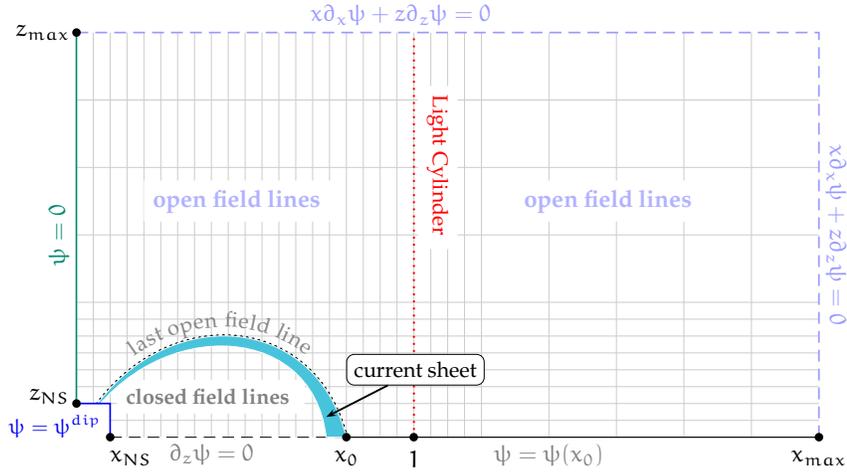}
  \caption{Calculation domain and boundary conditions}
  \label{fig:calc}
\end{figure}

We consider the force-free magnetosphere of an aligned rotator.  Such
magnetosphere is well described by the so-called pulsar equation
\citep{Michel73,ScharlemannWagoner73,Okamoto74}. For young pulsars the
differential rotation of the open field lines could be neglected and
the pulsar equation takes the form \citep[see e.g.][]{Goodwin/04}
\begin{equation}
  \label{eq:PsrEq_Om}
  (x^2-1)(\pd_{xx}\psi + \pd_{zz}\psi) +
  \frac{x^2+1}{x} \pd_x \psi -
  S \frac{d S}{d \psi}  =  0  
\end{equation}
All coordinates are normalized to the light cylinder radius
$\RLC\equiv c/\Omega$.  We use normalization for $\psi$, where the
dipole magnetic field function near the NS surface is given by
\begin{equation}
  \label{eq:PsiDipol}
  \psi^\mathrm{dip} = \frac{x^2}{(x^2+z^2)^{3/2}},
\end{equation}
The normalized poloidal current function $S$ is given by $S\equiv
(4\pi/c) (\RLC^2/\mu) I$, where $\mu\equiv B_0 \RS^3/2$ is the
magnetic momentum of the NS.  Magnetic field is expressed through the
functions $I$ and $\Psi\equiv(\mu/\RLC)\psi$ as
\begin{equation}
  \B = \frac{ \vec{\nabla} \Psi \times \vec{e_\phi} }{ \varpi } +
  \frac{4\pi}{c} \frac{I}{\varpi} \vec{e_\phi}
  \label{eq:B}
\end{equation}
The poloidal current $S$ is found from the condition at the light
cylinder
\begin{equation}
  \label{eq:LC_cond}
  \left. \pd_x \psi \right|_\mathrm{LC} =
  \frac1{2} \left.  S S^\prime \right|_\mathrm{LC}
\end{equation}
We assume existence of an equatorial current sheet with the return
current. The equatorial current sheet can be excluded from the
numerical treatment by setting appropriate boundary conditions at the
equatorial plane. For the current sheet at the boundary between closed
and open field line regions this is not possible, and we smear the
return current flowing along the last closed magnetic field line over
the region [$\psi_\mathrm{last}-d\psi,\psi_\mathrm{last}$], see
Fig.~\ref{fig:calc}.

A solution of the pulsar equation~(\ref{eq:PsrEq_Om}) for dipolar
magnetic field geometry had been found for the first time by
\cite{CKF}. They assumed, that the last closed field line is fixed at
the light cylinder. Later \cite{Goodwin/04} have found solutions for
different positions $x_0$ of the point, where the last closed field
line intersects the equatorial plane, inside the light cylinder,
$x_0\le1$. However they have made a rather unnatural assumption, that
the plasma pressure inside the closed field line domain is finite and
plays an important role in the force balance across magnetic field
lines.

We perform calculations for different $x_0\le 1$ with high numerical
resolution using grids with number of points in each directions
ranging from 2000 to 6000. In contrast to~\cite{Goodwin/04} we assume
zero pressure inside the closed field line region, i.e. the plasma
there is cold.  Equation~(\ref{eq:PsrEq_Om}) is solved numerically in
the domain $x_\mathrm{NS}\le x \le x_\mathrm{max},z_\mathrm{NS}\le z
\le z_\mathrm{max}$.  Boundary conditions are shown in
Fig.~\ref{fig:calc}.  The equation~(\ref{eq:PsrEq_Om}) is solved by a
full multigrid (V-cycles) FAS scheme \citep[see][]{TrottenbergBook}.
We use Gauss-Seidel smoother and SOR Gauss-Seidel solver at the
coarsest level.  Independence of the obtained results on the domain
sizes ($x_\mathrm{NS},x_\mathrm{max},z_\mathrm{NS},z_\mathrm{max}$),
$d\psi$ and algorithm-specific parameters has been verified.

We have expanded $\psi$ at the light cylinder in the Taylor series up
to the second order terms, and substituted this expansion into the
pulsar equation imposing the continuity of $\psi$ at the light
cylinder. In this way we obtained an equation for $\psi$ \emph{at} the
light cylinder, accurate up to the second order term, what is
sufficient for our numerical treatment, because we use a second order
scheme. Eq.(\ref{eq:LC_cond}) is used for determination of the
poloidal current. Such techniques was used by \cite{Goodwin/04} in
their simulations. It allows numerical treatment of the problem in a
single domain, rather than doing extremely CPU time consuming matching
of the solutions inside an outside the light cylinder, as it was made
by \cite{CKF}.

\section{Main Results}

\begin{figure}
  \includegraphics[width=\textwidth]{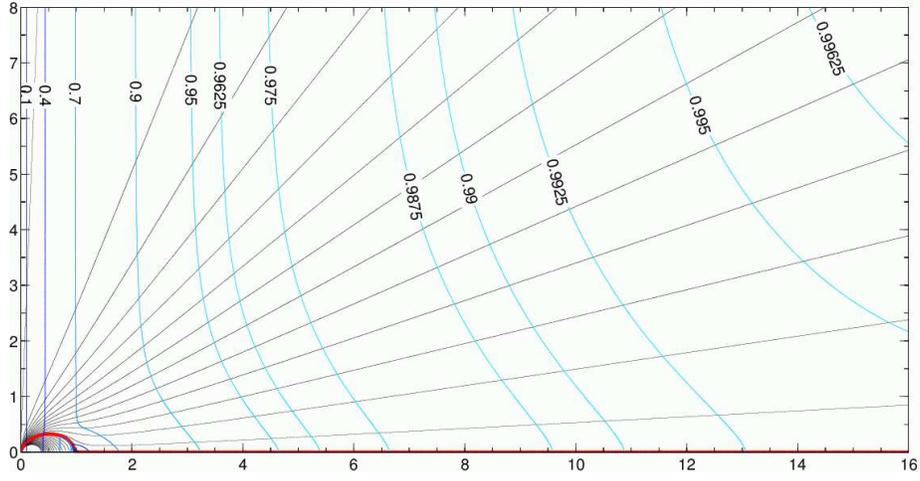}
  \caption{Magnetic field lines configuration (radial black lines) for
    $x_0=0.99, d\psi=0.08$. The last closed field line is shown by the
    thick red line.  Contours of the absolute values of the drift velocity
    normalized to the speed of light, $|\vec{u_\mathrm{D}}|/c$, are
    shown by the blue vertical lines.}
  \label{fig:x99g}
\end{figure}

\begin{figure}
  \includegraphics[width=\textwidth]{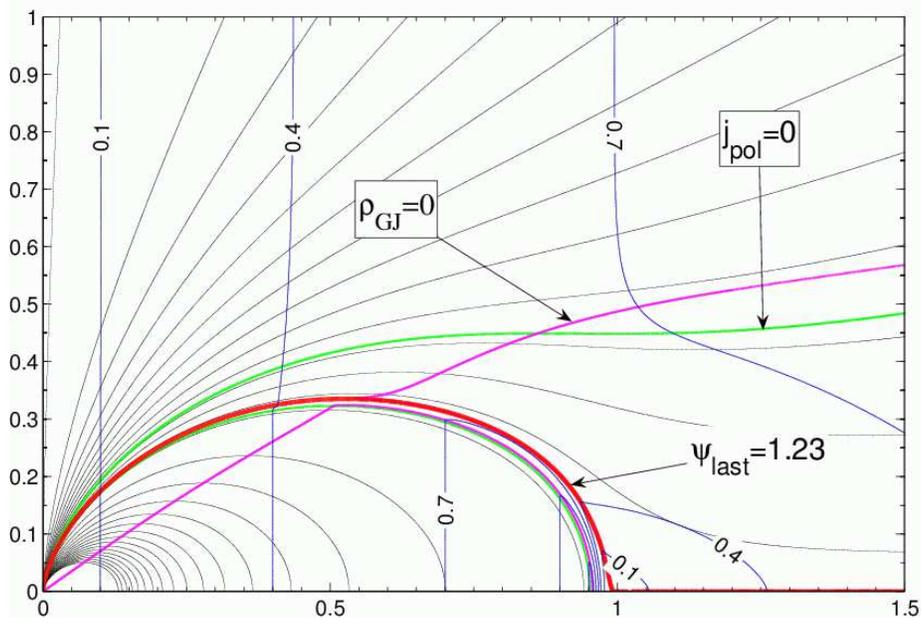}
  \caption{Central parts of the magnetosphere for
    $x_0=0.992,d\psi=0.04$. Notations similar to ones in
    Fig.~\ref{fig:x99g}.  Also line where $\Rgj$ changes sign and
    $j_{pol}=0$ are shown.}
  \label{fig:x99c}
\end{figure}

\begin{figure}
  \includegraphics[width=\textwidth]{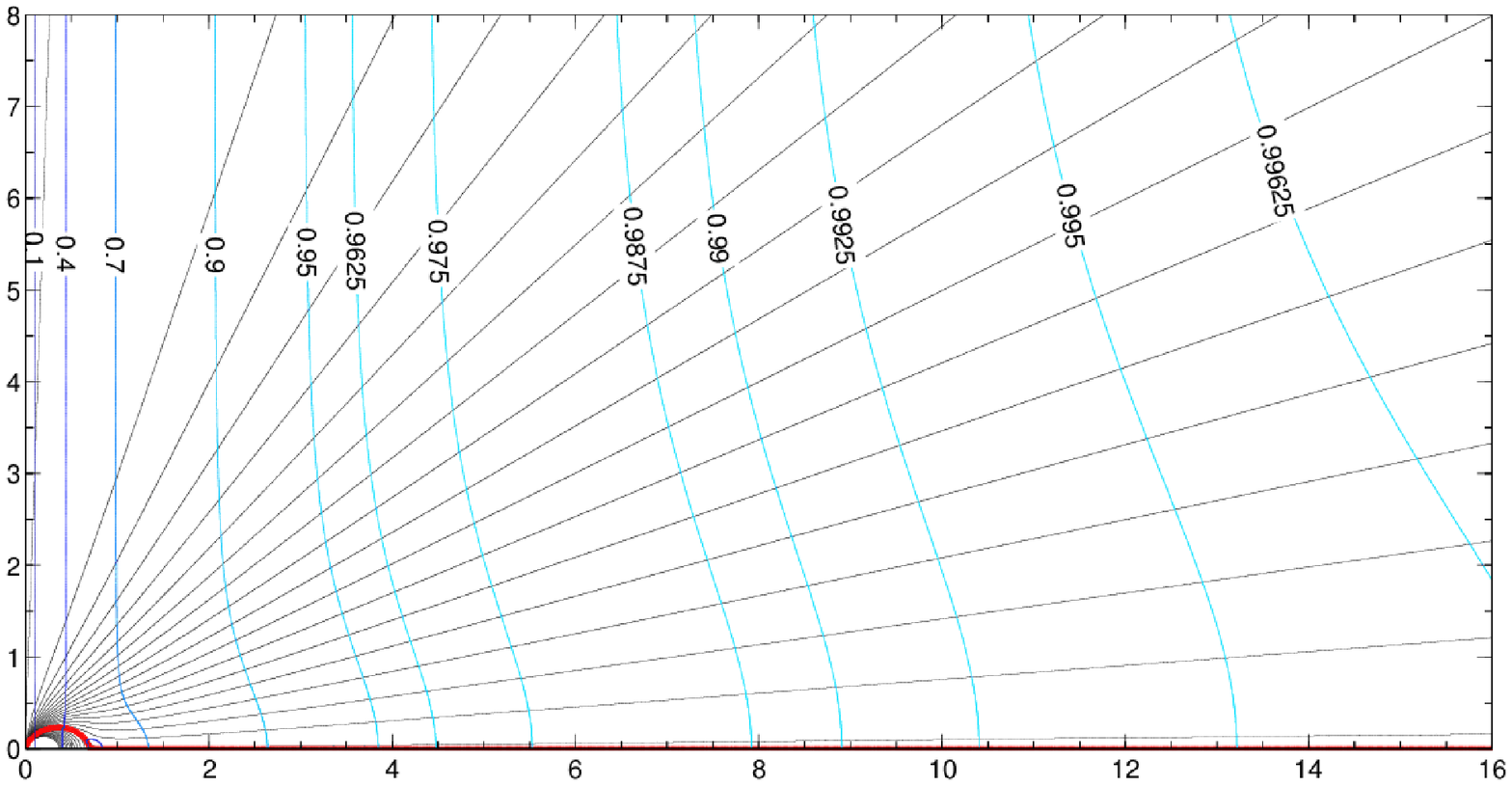}
  \caption{Drift velocity and magnetic field
    configuration for $x_0=0.7, d\psi=0.03$}
  \label{fig:x7g}
\end{figure}

\begin{figure}
  \includegraphics[width=\textwidth]{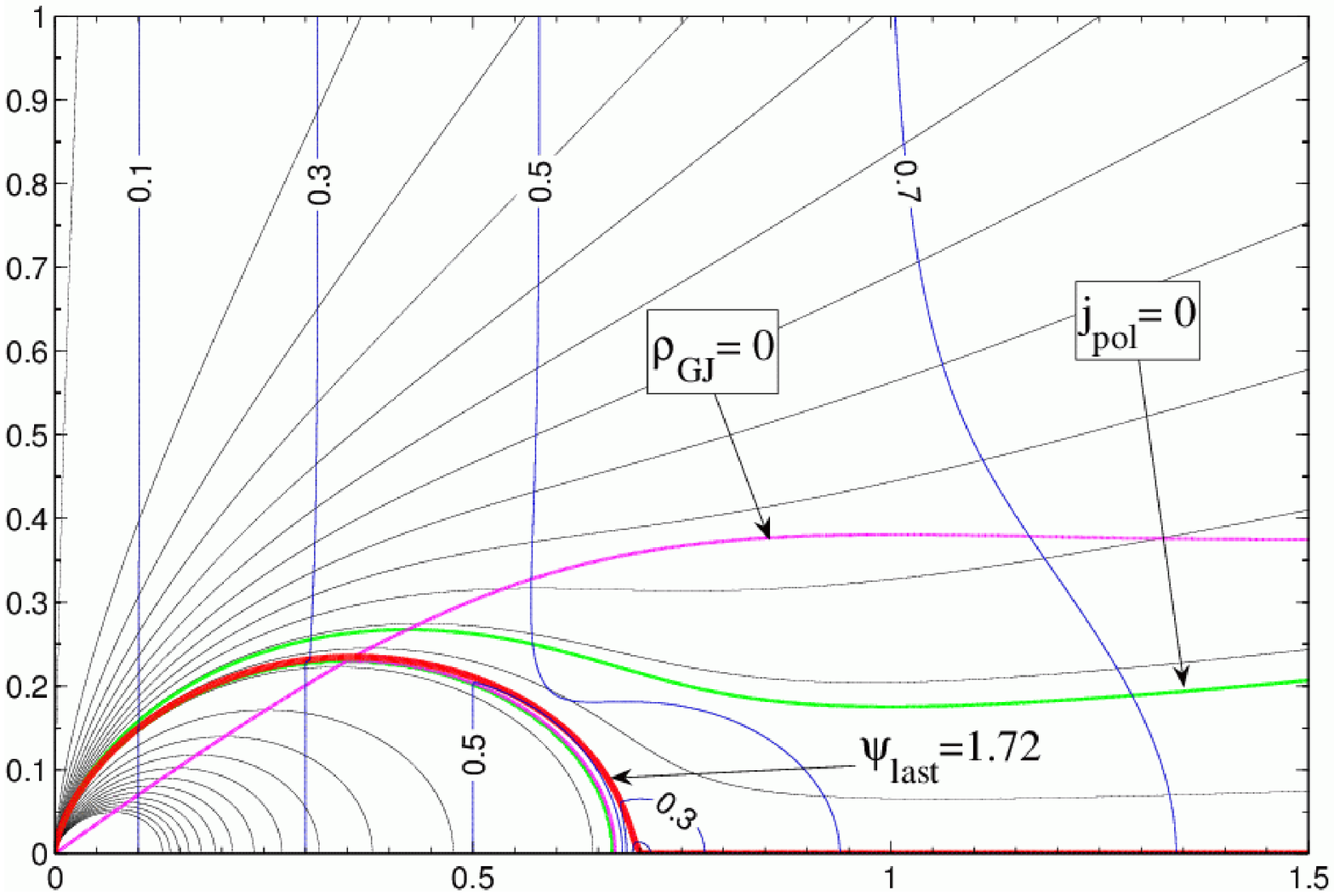}
  \caption{Inner parts of Fig.~\ref{fig:x7g}.}
  \label{fig:x7c}
\end{figure}

\begin{figure}
  \includegraphics[width=\textwidth]{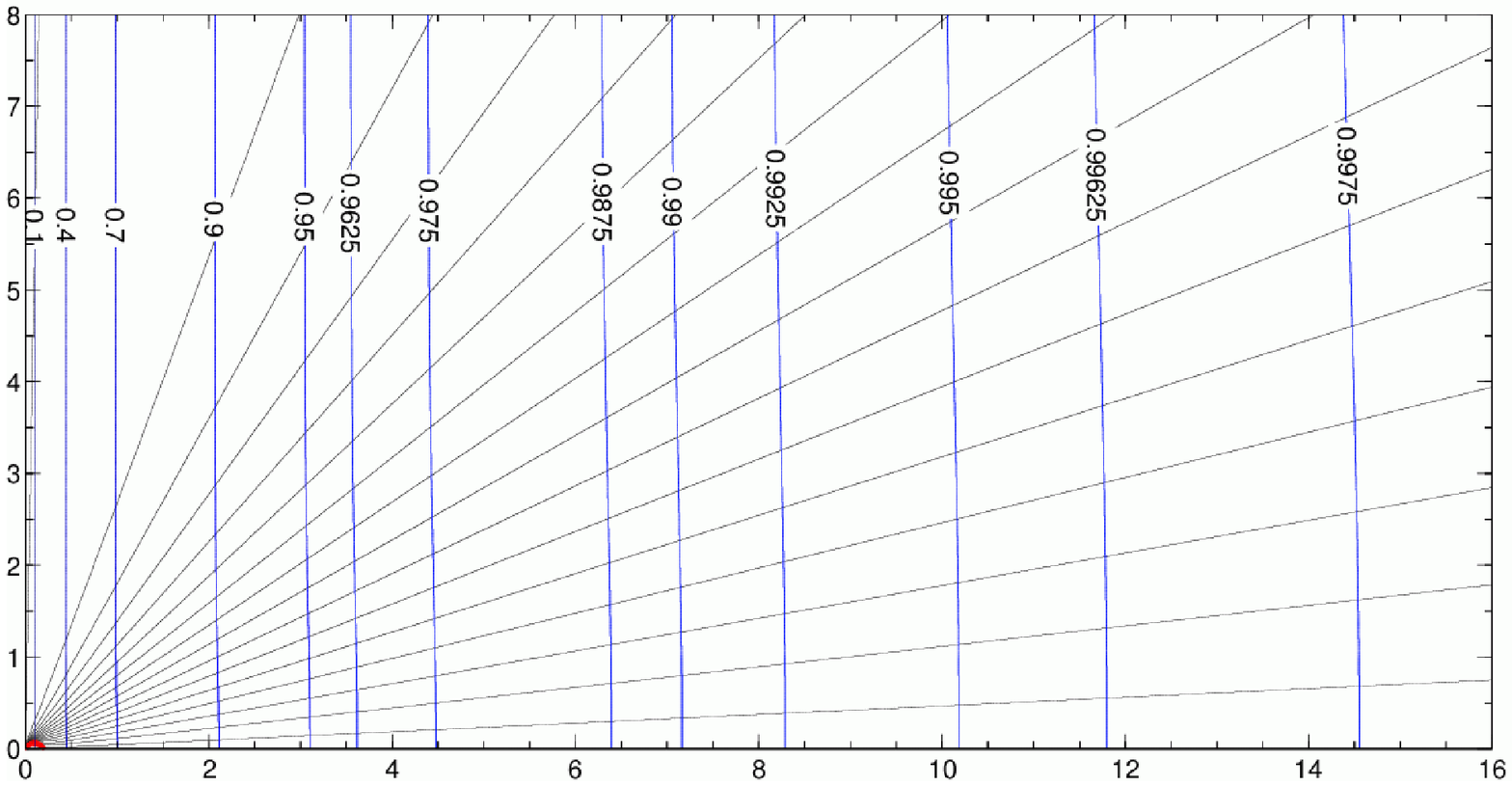}
  \caption{Drift velocity and magnetic field
    configuration for $x_0=0.2, d\psi=0.03$}
  \label{fig:x2g}
\end{figure}

\begin{figure}
  \includegraphics[width=\textwidth]{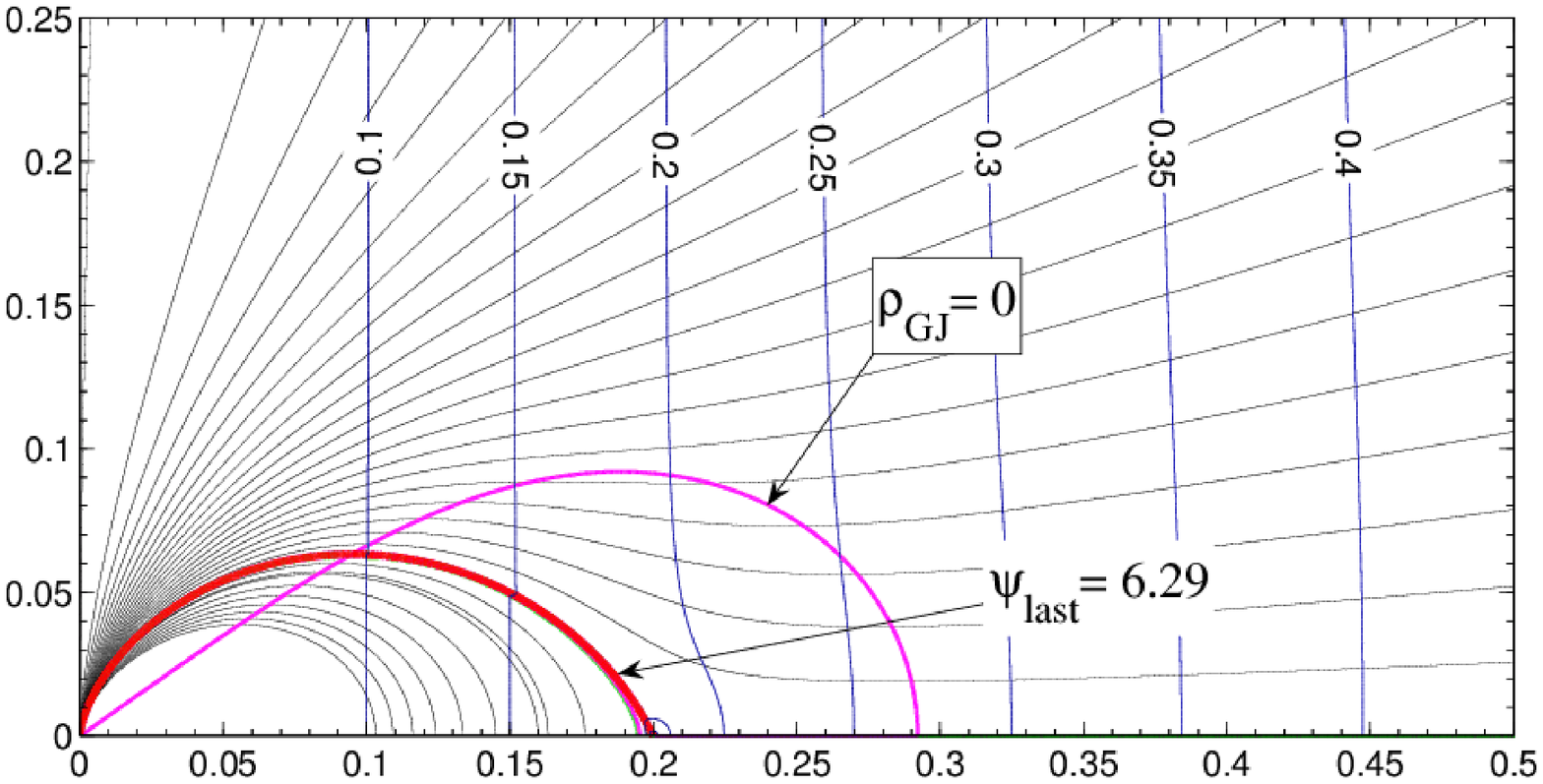}
  \caption{Inner parts of Fig.~\ref{fig:x2g}.}
  \label{fig:x2c}
\end{figure}

\begin{figure}
  \includegraphics[width=0.85\textwidth]{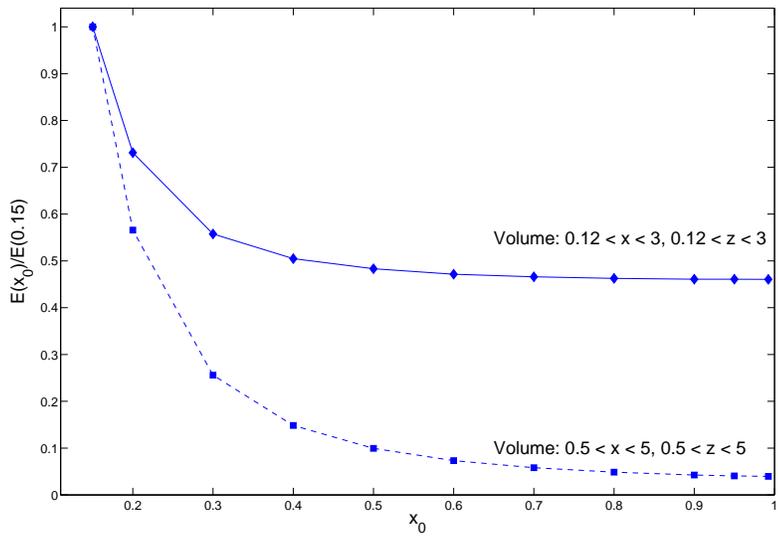}
  \caption{Total electromagnetic energy of two different 
    volumes as a function of $x_0$. The values were normalized to the
    energy in the corresponding volumes for $x_0=0.15$.}
  \label{fig:W_E}
\end{figure}

\begin{figure}
  \includegraphics[width=\textwidth]{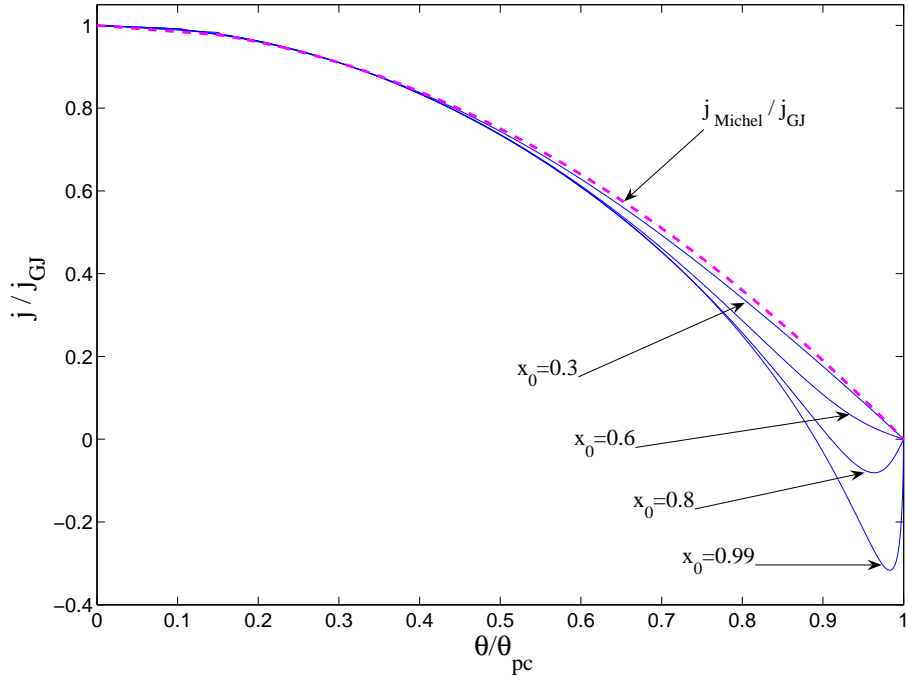}
  \caption{Poloidal current density in the polar cap of pulsar for
    different values of $x_0$, normalized to the corresponding
    Goldreich-Julian current density as a function of the colatitude,
    normalized to the colatitude of the polar cap boundary
    $\theta_\mathrm{pc}$. Michel current density is shown by the
    dashed line}
  \label{fig:jpc}
\end{figure}

\begin{itemize}

\item Calculations have been performed for the following values of
  $x_0$: 0.15; 0.2; 0.3; 0.4; 0.5; 0.6; 0.7; 0.8; 0.9; 0.95; 0.992.
  An unique solution has been found for each of the above $x_0$'s.  As
  representative examples, in Figs.~\ref{fig:x99g}-\ref{fig:x2c} properties
  of solutions for three cases ($x_0=0.2; 0.7; 0.99$) are shown.
  
\item Each solution has been checked for applicability of the
  force-free condition $E<B$. In none of them this condition is
  violated, at least up to 16 light cylinder radii from the NS. 
  The force-free condition can be reformulated as the
  condition on the drift velocity
  \begin{equation}
    \vec{u_\mathrm{D}} = \frac1{c} \frac{\E \x \B }{B^2} < c.
    \label{eq:Udrift:general}
  \end{equation}
  In Figs.~\ref{fig:x99g},\ref{fig:x7g},\ref{fig:x2g} maps of
  $u_\mathrm{D}$ for three values of $x_0$:0.2,0.6 and .99, are
  shown. One can see, that the critical value of $u_\mathrm{D}=1$ is
  never achieved. In all other cases the situation is similar.

\item For $x_0$ close to 1 ($x_0=0.95; 0.99; 0.992$) the magnetic
  field strength $B$ at the equatorial plane increases very rapidly
  when $x$ approaches $x_0$. The maximum value of $B$, achieved in the
  closed field line domain near the point $x_0$, increases with
  increasing of $x_0$ and decreasing of $d\psi$, see
  Fig.~\ref{fig:calc}. We interpret this as a result of
  $\lim_{x\rightarrow 1}B=\infty$, predicted by~\cite{Gruzinov:PSR}.

\item Energy losses of the pulsar for each $x_0$ have been calculated numerically.
  They increase with decreasing of $x_0$ and can be described by the formula
  \begin{equation}
    \label{eq:dWdt}
    W = A(x_0)*W_{md} \equiv  A(x_0)*\frac{2}{3}\frac{\mu^2\Omega^4}{c^3}, 
  \end{equation}
  where $W_{md}$ are magnetodipolar energy losses.
  The function $A(x_0)$ can be fitted surprisingly well by the power law
  \begin{equation}
    \label{eq:A_x0}
    A(x_0) = 1.41 \; x_0^{-2.065} \, , 
  \end{equation}
  see Fig.~\ref{fig:W_E}. We note, that analytical estimations of the
  energy losses based on the Michel poloidal current distribution for
  the dipolar magnetic field gives
  \begin{equation}
    \label{eq:A_x0_analyt}
    A(x_0) = x_0^{-2} 
  \end{equation}
  
\item The total energy of the electromagnetic field in the
  magnetosphere $E_\mathrm{tot}=\int \frac{B^2+E^2}{8\pi}\, dV$ decreases
  with increasing of $x_0$, see Fig.~\ref{fig:W_E}.  

\item In configurations with $x_0>0.6$ there is a
  \emph{volume} return current flowing along the open magnetic field
  lines, which carries however only a small part of the whole return
  current. For $x_0\le 0.6$ the return current flows only along the
  last open field line. On Fig.\ref{fig:jpc} the volume current density
  along open field lines in the polar cap of pulsar is shown. One can see,
  that the current density (in units of $j_{GJ}$) close to the polar
  cap boundary increases with increasing of $x_0$. The current
  density does not exceed the Goldreich-Julian current density
  $\GJ{j}\equiv\Rgj c$ and at most field lines it is
  \emph{less} than $\GJ{j}$

\end{itemize}

\section{Evolution of the pulsar magnetosphere}
For each $x_0$ a special unique volume current distribution is
necessary in order to support a force-free configuration of the
magnetosphere. The current density along open field lines in the polar
cap of pulsar could adjust to the values required by the global
magnetosphere structure (both smaller and greater than $\GJ{j}$) when
there is a particle \emph{inflow} from the magnetosphere into the
polar cap acceleration zone \citep{Lyubar92}.  As a source of
particles in the magnetosphere the outer gap cascade could be
considered. On the other hand, when the polar cap cascade is strong
and produces a lot of secondary particles with a wide energy
distribution, some of these particles with the smallest
energies, can be reversed in the outer magnetosphere either as a
result of momentum redistribution in the turbulent outflowing plasma,
or by a weak electric field arising as the magnetosphere will try to
achieve the most energetically favorably, i.e. force-free,
configuration.  The polar cap and the outer gap cascades can not
supply any arbitrary particle density, and, hence, can not support any
arbitrary current density in the magnetosphere. However, a current
with density greater or smaller than \GJ{j} can flow through the polar cap
acceleration zone only if there is a particle inflow from the
magnetosphere. The large the deviation of $j$ from \GJ{j}, the larger
particle flux from the magnetosphere is necessary. So, when the pulsar
becomes older, the range of current densities, which could be
supported by the cascades becomes smaller and the magnetosphere should
change configuration to a new one, where the required current
density could be supported by the weaker cascades.

As the electromagnetic energy of the magnetosphere decreases with
increasing of $x_0$, the most preferably configuration is that, where
the null point (point where the last closed field line intersects the
equatorial plane) lies at the light cylinder. However, in such
configuration the deviation of the  poloidal current density at the
open field lines from the Goldreich-Julian current density is
the largest and, hence, this requires a powerful source of the
particles in the magnetosphere. When the pulsar becomes older and
cannot support such poloidal current, the null point moves to a new
position, closer to the NS, corresponding to a magnetosphere
configuration, where the deviation of the current density from \GJ{j} is
smaller and such current density could be supported by the weaker
cascades. In the new configuration the energy losses (for the same
$\Omega$) are larger than in the configuration with $x_0=1$ and, hence,
the energy losses of pulsar decrease with time slower than it is given
by the magnetodipolar formula. If at some time the dependence of $x_0$
on the pulsar age and, hence, on $\Omega$ is given by $x_0\propto
\Omega^\xi$, then using
eqs.~(\ref{eq:A_x0}),(\ref{eq:dWdt}) we get
\begin{equation}
  \label{eq:dwdt_final}
  W \propto \Omega^\alpha,\qquad \alpha=4-2.065\,\xi,     
\end{equation}
$\xi$ is a complicated function of the pulsar age $t$, but as $x_0$
decreases with $t$ (see above), it is always positive, $\xi(t)>0$.
So, the braking index 
\begin{equation}
  \label{eq:n_braking}
  n = \frac{\ddot{\Omega}\Omega}{\dot{\Omega}^2} = \alpha-1 = 3-2.065\,\xi,     
\end{equation}
is always less than 3!. 

Detailed description of the numerical method and results of
calculation, as well as discussion of their implication for pulsar
physics, are given in \cite{Timokhin05}.

\noindent This work was supported by RFBR grant 04-02-16720.

\bibliographystyle{mn2e}
\bibliography{/home/atim/ARTICLES/PSREQ_1/draft/psreq}

\begin{thebibliography}{}

\bibitem[\protect\citeauthoryear{{Contopoulos}, {Kazanas} \&
  {Fendt}}{{Contopoulos} et~al.}{1999}]{CKF}
{Contopoulos} I.,  {Kazanas} D.,    {Fendt} C.,  1999, \apj, 511, 351

\bibitem[\protect\citeauthoryear{{Goodwin}, {Mestel}, {Mestel} \&
  {Wright}}{{Goodwin} et~al.}{2004}]{Goodwin/04}
{Goodwin} S.~P.,  {Mestel} J.,  {Mestel} L.,    {Wright} G.~A.~E.,  2004,
  \mnras, 349, 213

\bibitem[\protect\citeauthoryear{Gruzinov}{Gruzinov}{2005}]{Gruzinov:PSR}
Gruzinov A.,  2005, Phys.Rev.Lett., 94, 021101

\bibitem[\protect\citeauthoryear{{Lyubarskij}}{{Lyubarskij}}{1992}]{Lyubar92}
{Lyubarskij} Y.~E.,  1992, \aap, 261, 544

\bibitem[\protect\citeauthoryear{{Michel}}{{Michel}}{1973}]{Michel73}
{Michel} F.~C.,  1973, \apjl, 180, L133

\bibitem[\protect\citeauthoryear{{Okamoto}}{{Okamoto}}{1974}]{Okamoto74}
{Okamoto} I.,  1974, \mnras, 167, 457

\bibitem[\protect\citeauthoryear{{Scharlemann} \& {Wagoner}}{{Scharlemann} \&
  {Wagoner}}{1973}]{ScharlemannWagoner73}
{Scharlemann} E.~T.,  {Wagoner} R.~V.,  1973, \apj, 182, 951

\bibitem[\protect\citeauthoryear{Timokhin}{Timokhin}{2005}]{Timokhin05}
Timokhin A.~N.,  2005, in preparation

\bibitem[\protect\citeauthoryear{{Trottenberg}, {Oosterlee}, {Sch\"uller},
  {Brandt}, {Oswald} \& {St\"uben}}{{Trottenberg}
  et~al.}{2001}]{TrottenbergBook}
{Trottenberg} U.,  {Oosterlee} C.~W.,  {Sch\"uller} A.,  {Brandt} A.,  {Oswald}
  P.,    {St\"uben} K.,  2001, Multigrid.
Academic Press

\end{thebibliography}

\newpage

\begin{figure}
  \includegraphics[width=\textwidth]{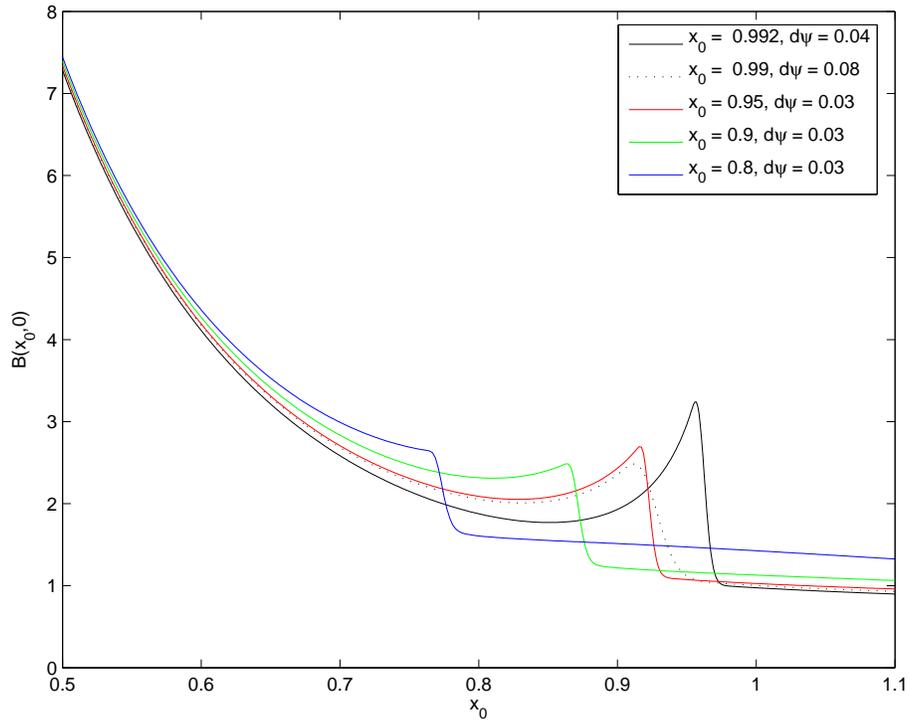}
  \caption{Magnetic field strength at the equatorial plane as a function
    of $x$ for different $x_0$ and widths of the current sheet.}
  \label{fig:B_x}
\end{figure}

\begin{figure}
  \includegraphics[width=0.85\textwidth]{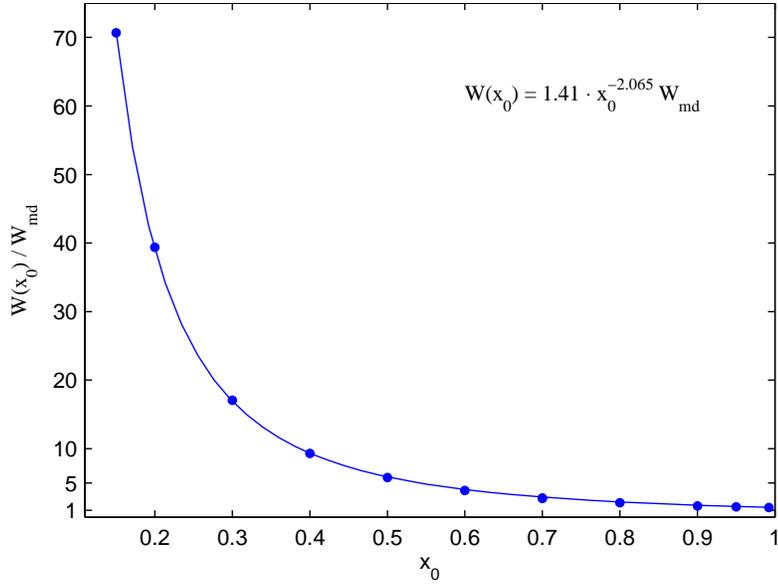}
  \caption{Numerically calculated energy losses normalized to the
    magnetodipolar energy losses as a function of $x_0$ (dots). Fit of
    the results by the power law is shown by the solid blue line.}
  \label{fig:W}
\end{figure}

\end{document}